\documentclass[12pt,preprint]{aastex}

\shorttitle{IR observations of 1E 161348-5055}
\shortauthors{A. De Luca et al.}

\newcommand{\chan}{{\sl Chandra}}
\newcommand{\rosat}{{\sl ROSAT}}
\newcommand{\asca}{{\sl ASCA}}

\newcommand{\xmm}{{\sl XMM-Newton}}

\newcommand{\hst}{{\sl HST}}
\newcommand{\vlt}{{\sl VLT}}
\newcommand{\vltn}{{\sl Very Large Telescope}}

\newcommand{\fors}{{\sl FORS1}}
\newcommand{\naco}{{\sl NACO}}
\newcommand{\isaac}{{\sl ISAAC}}
\newcommand{\nacon}{{\sl NAos COnica}}
\newcommand{\nicmos}{{\sl NICMOS}}
\newcommand{\ein}{{\sl Einstein}}

\newcommand{\tmass}{{\sl 2MASS}}
\newcommand{\gsc}{{\sl GSC-2}}

\newcommand{\dao}{{\em Daophot}}

\newcommand{\all}{{\em Allframe}}

\begin{document}


\title{Deep infrared observations of 
the puzzling central X-ray source in RCW103\altaffilmark{2,3}}


\author{A. De Luca\altaffilmark{1}}
\affil{INAF - Istituto di Astrofisica Spaziale e Fisica Cosmica,
Via Bassini 15, I-20133 Milano, Italy}
\email{deluca@iasf-milano.inaf.it}
\and
\author{R.P. Mignani}
\affil{University College London, Mullard Space Science Laboratory, Holmbury St. Mary, Dorking, Surrey, RH5 6NT Unite
d Kingdom}
\and
\author{S. Zaggia}
\affil{INAF - Osservatorio Astronomico di Padova, Vicolo dell'Osservatorio 5,
  I-35122, Padova, Italy}
\and
\author{G. Beccari}
\affil{INAF - Osservatorio Astronomico di Bologna, Via Ranzani 1, I-40127,
  Bologna, Italy }
\and
\author{S.Mereghetti, P.A. Caraveo\altaffilmark{2}}
\affil{INAF - Istituto di Astrofisica Spaziale e Fisica Cosmica,
Via Bassini 15, I-20133 Milano, Italy}
\and
\author{G.F. Bignami\altaffilmark{1}}
\affil{Agenzia Spaziale Italiana, Via Liegi 26, I-00198 Roma, Italy}
\altaffiltext{1}{Istituto Universitario di Studi Superiori di Pavia, V.le
  Lungo Ticino Sforza 56, 27100
  Pavia, Italy}

\altaffiltext{2}{Based on observations 
collected at the European Southern Observatory, Paranal, Chile under
programme ID 67.D-0198(A), 
077.D-0764(A)} 
\altaffiltext{3}{Based on observations with the NASA/ESA 
Hubble Space Telescope, obtained at the Space Telescope Science 
Institute, which is operated by AURA, Inc. under contract No. 
NAS 5-26555.}



\begin{abstract}
1E 161348-5055 (1E 1613) is a point-like, soft X-ray source 
originally identified as a radio-quiet, isolated neutron star,
shining at the center of
the 2000 yr old supernova remnant RCW103.
 1E 1613 features a puzzling 6.67 hour periodicity as well as a
dramatic variability 
over a time scale of few years.
Such a temporal behavior,
coupled to the young age and to the lack of an obvious optical counterpart,
makes  1E 1613 a unique source among all compact objects associated to SNRs.
It could either be the first low-mass X-ray binary system discovered
inside a SNR,
or a peculiar isolated magnetar with an extremely slow 
spin period. 
Analysis of archival IR observations, performed in 2001 with the VLT/ISAAC
instrument, and in 2002 with the NICMOS camera onboard HST unveils
a very crowded field. A few sources are positionally consistent with the
refined X-ray error region that we derived from the analysis of 13 Chandra observations.
To shed light on the nature of 1E 1613, we have performed 
deep IR observations of the field with the NACO instrument at the
ESO/VLT, 
searching for variability. 
None of the candidates, however, shows a clear modulation 
at 6.67 hours, nor has a significant long term variability. 
Moreover, none of the candidates stands out for peculiar
colors  with  respect  to  the  bulk of the sources detected in the field.  
We find no compelling reasons to associate any of the candidates to 1E 1613. 
 On one side, within the frame of the binary system model 
 for the X-ray source, it is very unlikely that one of the candidates 
 be a low-mass companion star to 1E 1613. On the other side,  
 if the X-ray source is an isolated magnetar surrounded by  
 a fallback disc, we cannot exclude that the IR counterpart be hidden 
 among the candidates. 
 If none of the potential counterparts is linked to the X-ray source, 
 1E 1613 would remain undetected in the IR down to
 Ks$>$22.1. 
Such an upper limit is consistent only with an extremely low-mass star 
(an M6-M8 dwarf) at the position of 1E 1613, and makes rather problematic the
interpretation of 1E 1613 as an accreting binary system. 
 \end{abstract}





\section{Introduction}

The  X-ray point  source 1E  161348$-$5055 (1E 1613 hereinafter) was
discovered with the \ein\  observatory close to the geometrical center
of  the very  young  ($\sim2000$ years)  shell-type supernova  remnant
(SNR) RCW103 \citep{tuohy80}.  The  association of the  point source
with  the SNR  is very  robust  on the  basis of  the good  positional
coincidence,  with  1E 1613     lying  within  $\sim20''$  of  the  SNR
center. Furthermore, HI  observations of this region \citep{reynoso04}
pointed to a  spatial correlation of the two objects  in view of their
similar distance ($\sim3.3$ kpc).  Historically, 1E 1613  was the first
radio-quiet,  isolated neutron star  candidate, 
with thermal
X-ray  spectrum, no  counterparts  at other
wavelengths, no  pulsations, no non-thermal  extended emission.  Since
then, a handful of similar  enigmatic sources, all characterized by a
thermal
X-ray    spectrum, lack of    standard     pulsar    activity,
high X-ray to optical flux ratio 
(F$_X$/F$_{opt}>10^3$),  and  general  lack  of  pulsations,  have  been
discovered  inside young  SNRs.  Such  sources, possibly  the youngest
members  of   the  family  of  radio-quiet,   isolated  neutron  stars
(including also the Anomalous X-ray Pulsars and Soft Gamma Repeaters),
were    dubbed,   as    a   class,    ``Central    Compact   Objects''
\citep[CCOs,][]{pavlov02} -- see \citet{deluca07} for a recent review.

What makes 1E 1613  unique among CCOs is its very peculiar and puzzling
temporal behaviour.   A factor  10 variability on  the time
scale of few years  was already  evident within  the  historical \ein/\rosat/\asca\
dataset   \citep{gotthelf99}.   This   was  confirmed   by  \chan/ACIS
observations  which showed  that  the source  brightened  by a  factor
$\sim$60  between  September   1999  
(when the source's flux was $\sim8 \times 10^{-13}$  erg  cm$^{-2}$
s$^{-1}$  in the 0.5-8  keV energy range)  and  February  2000  
(with a record flux of $\sim5 \times 10^{-11}$  erg
cm$^{-2}$  s$^{-1}$), 
decreasing
to  $\sim10^{-11}$ erg  cm$^{-2}$
s$^{-1}$  \citep{sanwal02}  afterwards.    The  analysis  of  \chan\
monitoring  observations   showed  that  the  source   flux  has  been
continuously  fading  since  then  \citep{deluca06}.  The  1999 16 ks
\chan\ observation, performed when the source was
in a ``low state'', hinted a possible $\sim$6.4 hours ($\sim$ 23 ksec)
periodicity \citep{garmire00}.   However, subsequent \chan\  and \xmm\
observations  found  the source  in  ``active  state''  
with a remarkably complex
light  curve  including dips, with  an  overall $\sim20\%$  modulation
\citep{sanwal02,becker02},
and could not
ultimately confirm its periodicity.  The  breakthrough came with a  long, 90 ks
\xmm\ observation \citep{deluca06}, performed when the source was in a
``low  state'' ($\sim2\times10^{-12}$  erg cm$^{-2}$  s$^{-1}$), which
clearly showed a 6.67 hour periodicity with a strong ($\sim50\%$), almost
sinusoidal modulation.

On the optical  side, \vlt\ observations of the  crowded field of 1E 1613
performed with \fors\ and \isaac\ identified a possible counterpart in
a very red object  (I$>25$, J$\sim22.3$, H$\sim19.6$ and Ks$\sim18.5$)
located  within   the  \chan\  error   circle  \citep{sanwal02}.   The
existence of this object was also confirmed by \hst/\nicmos\ follow-up
observations  \citep{mignani04}. 
A search for a counterpart
in the far IR with the Spitzer telescope was also performed,
with negative results \citep{wang07}. 

As   discussed   by
\citet{deluca06}, the  peculiar combination of  long-term variability,
6.67  hours  periodicity,   young  age  and  underluminous  optical/IR
counterpart settle the case for  a unique phenomenology. 1E 1613  could
be a  very young  binary system, composed  of a recently  born compact
object and  of a low-mass  star in an  eccentric orbit, powered  by an
unusual ``double'' (wind+disc) accretion mechanism.  
Very recently, a different binary scenario for 1E 1613 has been proposed
by \citet{pizzolato08}.
Alternatively, 1E 1613
 could be  a peculiar isolated object, e.g.  a   magnetar,  dramatically
slowed-down,  possibly  by  interaction   with  a  debris  disc
(De Luca et al., 2006; see also Li, 2007, for an update of the isolated magnetar 
model for 1E 1613).
Both
the binary system and the isolated object
scenario  are  highly unusual  and  require non-standard  assumptions
about  the formation  and evolution  of compact  objects  in supernova
explosions.

In order to  shed light on the nature of the  puzzling source 1E 1613 ,
we  have performed new,  deep IR  observations of  the field  with the
\vlt, with the main aim of identifying the source counterpart by means
of  a sensitive  search  for  modulation at  the  expected 6.67  hour
periodicity.  We  have also re-analysed  the archived \vlt\  and \hst\
observations  \citep{sanwal02,mignani04}   in  order  to   search  for
long-term variability of a possible counterpart. We presented a first
account of our VLT results in \citet{mignani07}. 

Since a precise position is of paramount importance for our counterpart search,
we have first reassessed 1E 1613 X-ray position using a set of Chandra archival
data together with a very recent deep Chandra observation performed by our group 
(\S 2). Using our new position, we have analysed our new IR VLT data, as 
well as the archival ones (\S 3 and 4). Results are discussed in \S 5.

\section{The X-ray position of 1E 1613}

In  order  to  maximize  the  identification  chances,
an   improved  X-ray   position   of   the  source   is
required. Although 1E 1613 has been extensively observed with \chan\,
almost all data 
collected
before  2002 are  of little  use to  derive the  source
position owing to  the presence  of offsets  in the  astrometry, as
computed  by  the  {\em Aspect  tool}
\footnote{http://cxc.harvard.edu/cal/ASPECT/fix\_offset/fix\_offset.cgi},  
and/or to source pile-up. Indeed,
we  found discrepancies  among the  reconstructed  target coordinates,
much larger than the  expected astrometric accuracy.  Thus, we decided
to  rely only  on  Chandra data  collected  after 2002,  for which  no
astrometric offsets are reported. Such  data include 12 short (4-5 ks)
observations  with  ACIS/S \citep{sanwal02,deluca06}  as  well as a  very
recent, deep  (80 ks) observation  performed with HRC/S by  our group.
The short  observations were performed with different roll
angles, with the source at a  large ($\sim8$ arcmin)
offaxis angle in order to  reduce pile-up effects.   
Data were  retrieved  from the  Chandra  X-ray Centre  (CXC)
Science Archive.   Calibrated (``level  2'') data were  produced using
the  Chandra  Interactive  Analysis  of  Observations  software  (CIAO
v.3.3).  The target position was  computed for each ACIS/S dataset by
performing a source detection in  the 0.5--10 keV range using the {\em
wavdetect} task.  After averaging the target coordinates computed from
each  dataset, we obtained  $\alpha (J2000)$=16$^h$  17$^m$ 36.228$^s$,
$\delta  (J2000)$=  -51$^\circ$  02'  24$\farcs7$ with  a  r.m.s.   of
0\farcs5   and   0\farcs25  in   right   ascension  and   declination,
respectively. The Chandra/ACIS astrometric accuracy for sources at 
offaxis angles larger than 3 arcmin is degraded with respect to the on-axis case
because of PSF blurring, but no systematic studies of
such an effect are available. Thus, we assume the observed r.m.s. on the
source coordinates as the 1$\sigma$ uncertainty on the X-ray position.
The target  position  in the  deep  HRC observation  was
computed  in  the  same  way, yielding  $\alpha(J2000)$=16$^h$  17$^m$
36.232$^s$,  $\delta  (J2000)$=  -51$^\circ$  02' 24$\farcs6$  with  a
nominal $1\sigma$   radial uncertainty of $0\farcs41$, according
to  the CXC calibration  
team\footnote{http://cxc.harvard.edu/cal/ASPECT/celmon/}. 
The HRC  coordinates are
perfectly  consistent with those  computed using  the 12  short ACIS/S
observations.  Combining  the two  measurements, we computed  the best
estimate  of  the source  coordinates,  i.e.  $\alpha  (J2000)$=16$^h$
17$^m$ 36.23$^s$, $\delta (J2000)$= -51$^\circ$ 02' 24$\farcs6$ with a
$1\sigma$  uncertainty   of  $0\farcs285$  and   $0\farcs185$  
in  right ascension and declination, respectively.

\section{IR Observations and data reduction}


\subsection{The 2006 VLT/NACO observations}

Our new set of IR observations was performed in visitor mode on May
23rd and  24th 2006  
with \nacon\  (\naco),  an adaptive  optics  (AO)  imager and  spectrometer
mounted at the  fourth Unit Telescope (UT4) of the ESO \vltn\ (\vlt) at  
the Paranal Observatory, Chile.  In order to
provide   the  best   combination  of   angular   resolution  and
sensitivity,  \naco\   was  operated  with  the  S27   camera 
giving a field of  view of $28''\times28''$ and a  pixel scale of
0\farcs027.  The only suitable  reference star for the adaptive optics
correction  was  the  \gsc\  star  S230213317483  ($V=15.2$),  located
21\farcs1 away from our target.   For this reason, the resulting image
quality  was  not  optimal  and  appeared to be
very  sensitive  to  the  atmospheric
conditions.  The Visual ($VIS$)  dichroic element and wavefront sensor
($4500-10000 \:  \AA$) were used.  Observations were  performed in the
$K_s$ ($\lambda=2.18 \: \mu$ ;  $\Delta \lambda= 0.35 \: \mu$) filter.

In  order to monitor  continuously the  potential counterpart
candidates  within  the \chan\  error
circle covering at  least two 6.67 hour cycles,
we  obtained  a  total  of  21 consecutive  observations  in  the  two
nights (see Table~\ref{NACOdata}). 
Each observation lasted about 2300 s and was split in
sequences  of  short randomly  dithered  exposures  with  Detector
Integration  Times (DIT) of 60 s.   The
airmass was mostly below 1.3,  while the seeing was rarely below $\sim
0\farcs8$, affecting  the  performance of  the  AO.   Sky
conditions  were photometric  in  both nights.   Night (twilight  flat
fields) and day time calibration frames (darks, lamp flat fields) were
taken daily  as part of  the \naco\ calibration plan.   Standard stars
from the  Persson et al.  (1998)  fields were observed  in both nights
for photometric  calibration.  The data have been  processed using the
ESO  \naco\
pipeline\footnote{www.eso.org/observing/dfo/quality/NACO/pipeline} and
the science images coadded.

\subsection{VLT/ISAAC archival data}

IR  observations of  1E 1613 were performed  in service  mode
between April  and July 2001
using  the  NIR  spectro-imager  \isaac\  mounted at  the  First  Unit
Telescope (UT1) of  \vlt. 
  The  Short Wavelenght  (SW) camera was  used, equipped  with a
Rockwell  Hawaii 1024$\times$1024  pixel Hg:Cd:Te  array, which  has a
projected  pixel   size  of  0\farcs148   and  a  field  of   view  of
152$''\times$152$''$. Observations  were performed  through  the $J$
($\lambda= 1.25 \mu$; $\Delta  \lambda= 0.29 \mu$), $H$($\lambda= 1.65
\mu$;  $\Delta \lambda=  0.30 \mu$)  and $K_s$  ($\lambda=  2.16 \mu$;
$\Delta \lambda= 0.27 \mu$) band  filters.  A total of 13 observations
were performed  in the  $H$-band with the  aim of pinpointing  the CCO
counterpart through the detection of a flux modulation, as suggested 
by the possible periodicity of the X-ray source hinted by the early 
Chandra observations  available at that epoch \citep{garmire00}.
Additional
pointings  in the  $J$ and  $K_s$-bands  were performed  to study  the
colors  of  the  candidate  counterpart (see  Table  \ref{ISAACdata}).

To  allow  for
subtraction  of the variable  IR sky  background each  observation was
split  in  sequences of  shorther  dithered  exposures (DIT=20  s in  the
$H$-band  and 40  s  in the others).  
The total integration times  per observation were 2000 s ($J$
and $K_s$ bands)  and 1000 s ($H$-band).  All  observations, 
with the exception of the July 23rd one, were taken
under photometric conditions,
with  a seeing  often  better  than 1\farcs0  and  airmass below  1.2.
Twilight flat fields, dark frames, as well as images of standard stars
from the  Persson et al.  (1998)  fields, were taken daily  as part of
the \isaac\  calibration plan.  The  data were reduced  and calibrated
using                  the                 ESO                 \isaac\
pipeline\footnote{http://www.eso.org/observing/dfo/quality/ISAAC/pipeline}. For
each exposure  sequence, single frames were registered  and coadded to
produce a background subtracted and cosmic-rays free image.

\subsection{HST/NICMOS archival data}

IR  observations of the field of 1E 1613 were  performed on
August 15th and October 8th 2002 with the \hst\
Observations were  performed with  \nicmos\ using  the NIC2
camera ($19\farcs2  \times 19\farcs2$ field of  view, 0\farcs075 pixel
size) with  the $110W$  ($\lambda = 1.128  \: \mu$, $\Delta  \lambda =
0.16 \:  \mu$), $160W$  ($\lambda = 1.606  \: \mu$, $\Delta  \lambda =
0.11 \: \mu$) and $205W$ ($\lambda  = 2.071 \: \mu$, $\Delta \lambda =
0.18 \: \mu$)  filters. To cope with visit  scheduling constraints the
target had  to be observed  for 10 spacecraft orbits  distributed over
two  different  visits.  
In each visit, a sequence
of six exposures  (2590 s each) was performed in  the $160W$ filter to
search   for  variability   of  the   originally   proposed  candidate
counterpart \citep{sanwal02}, and two  exposures in the $110W$ (935 s)
and  $205W$ (1007  s)  filters  to derive  color  information (see  Table
\ref{NICMOSdata}).     


To   decrease   the    instrumental   overheads,
observations were  taken in MULTIACCUM  mode, split in sequences  of 9
and  18  sub-exposures  in  the  $110W$  filter  and  in  the  others,
respectively.   The  data  were  downloaded from  the  European  \hst\
science      archive\footnote{http://www.stecf.org/archive/}     after
on-the-fly  recalibration  with the  best  reference files  available,
frame coaddition and cosmic ray filtering.

\subsection{\vlt\ and \hst\ astrometry}

In order to precisely register the \chan\ position on our IR images we
have  refined  the default  image  astrometry.  We  have computed  the
astrometric solution  on the \isaac\  images by fitting  the positions
and  coordinates  of 60  reference  stars  selected  from the  \tmass\
catalogue.  The reference  stars  positions have  been  computed by  a
two-dimensional  gaussian  fitting procedure  with  accuracies of  few
hundredths  of pixels.  The  astrometric fit  was performed  using the
STARLINK package ASTROM with a six order polynomial to account for the
detector distorsions and  yielded a rms of $\sim  0\farcs090$ in both
right ascension and declination. Since  very few \tmass\ stars fall in
both  the narrow  \nicmos\ ($19\farcs2  \times 19\farcs2$)  and \naco\
($28''\times28''$)  fields of  view they  do not  provide  an adequate
primary  reference  grid. For  this  reason,  in  both cases  we  have
computed the astrometric solution  by using a secondary reference grid
made up  of 26 secondary stars  identified in common  with the \isaac\
$Ks$-band image.  The astrometric solutions computed  for the \nicmos\
and \naco\ images thus yielded  a rms of $0\farcs042$ and $0\farcs040$
per coordinate, respectively.  By adding  in quadrature the rms of the
astrometric  solution  of  the  \isaac\ image  ($\sim  0\farcs090$  per
coordinate), we thus  end up with an overall  uncertainty of $\approx$
0\farcs1  per  coordinate  on  both  the \nicmos\  and  \naco\  images
astrometry.   In  all  cases,  we  have accounted  for  the  intrinsic
0\farcs2 absolute astrometric accuracy of \tmass
\footnote{http://spider.ipac.caltech.edu/staff/hlm/2mass/overv/overv.html}.

\begin{figure*}
\includegraphics[angle=0,width=7cm]{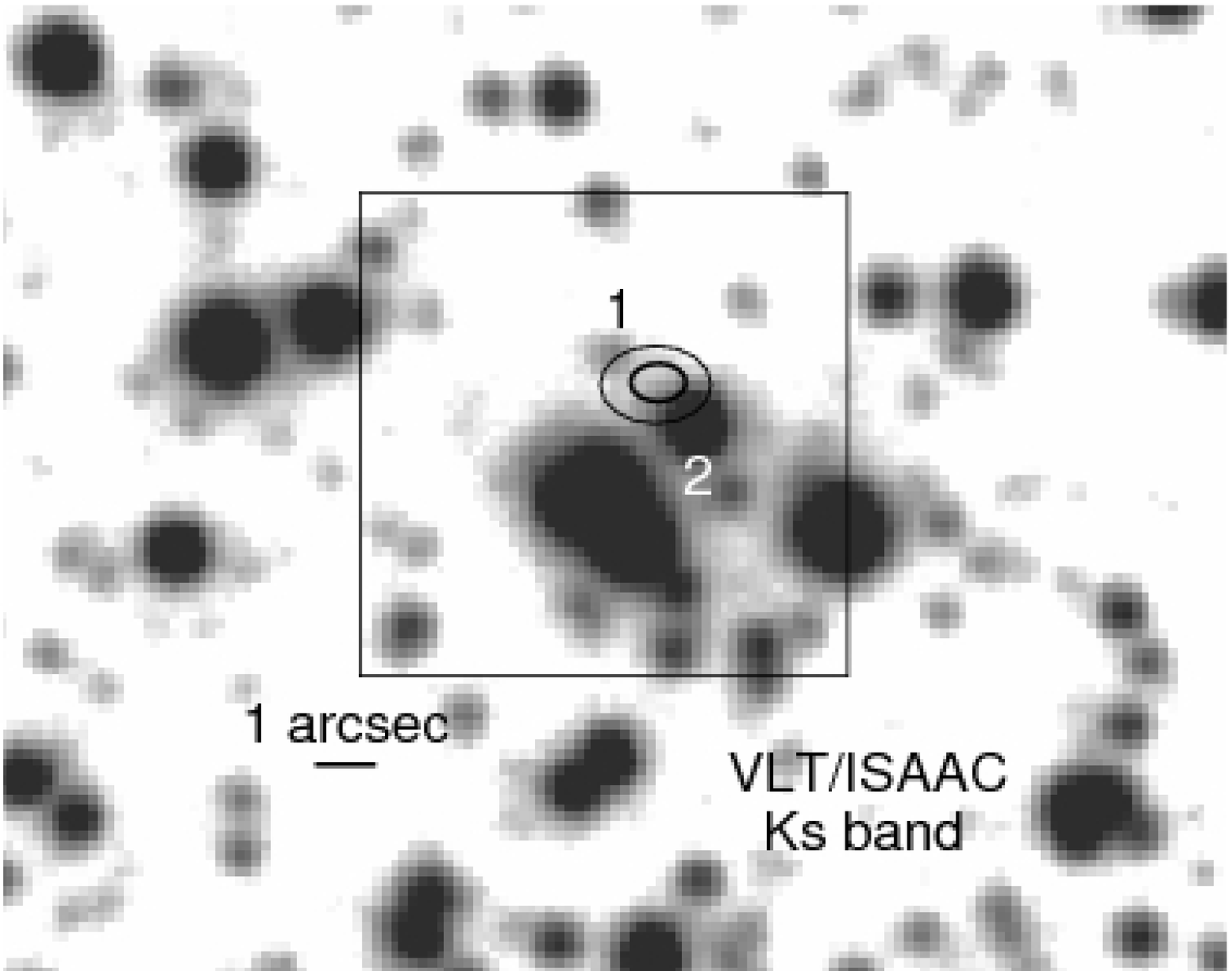} 
\includegraphics[angle=0,width=7cm]{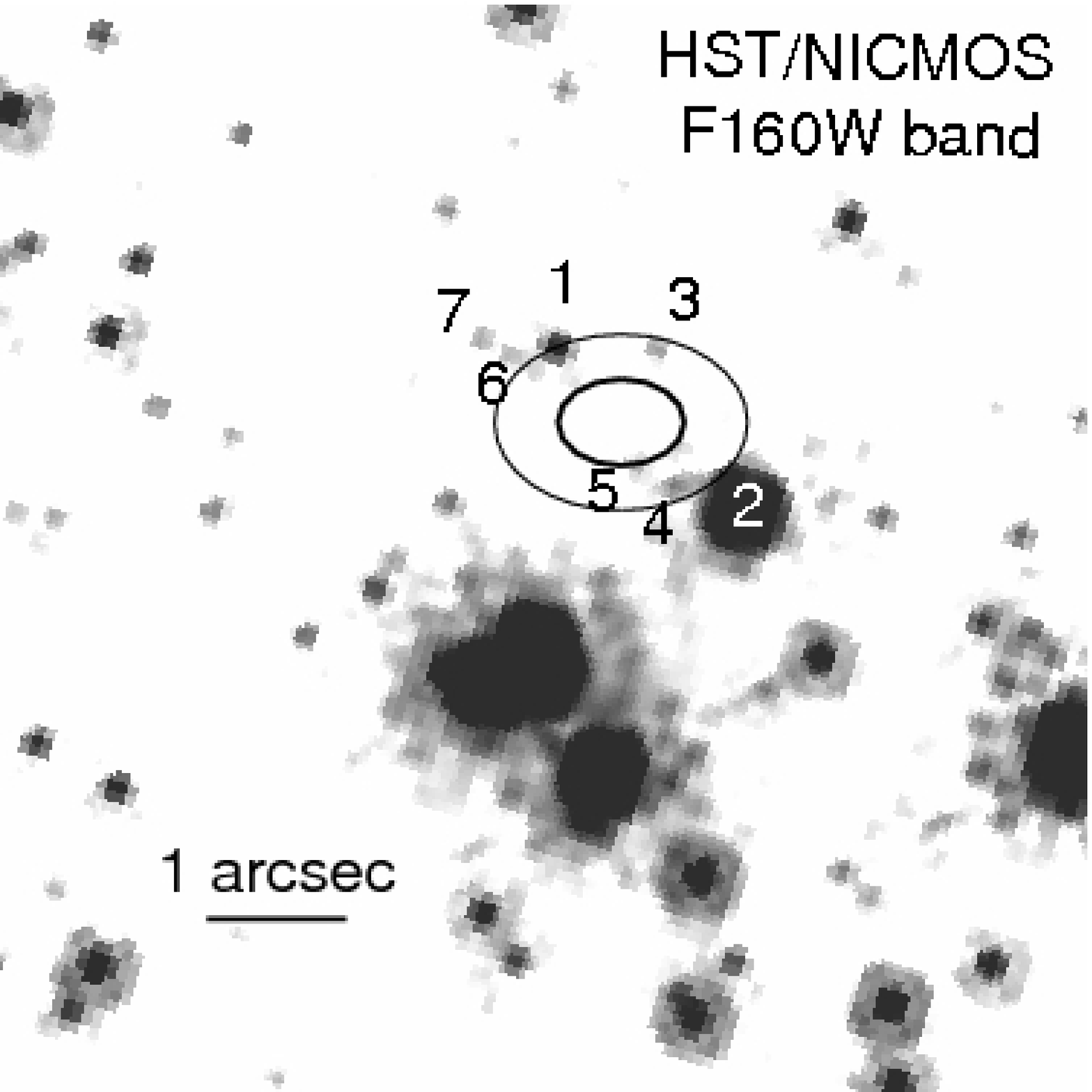}
\includegraphics[angle=0,width=7cm]{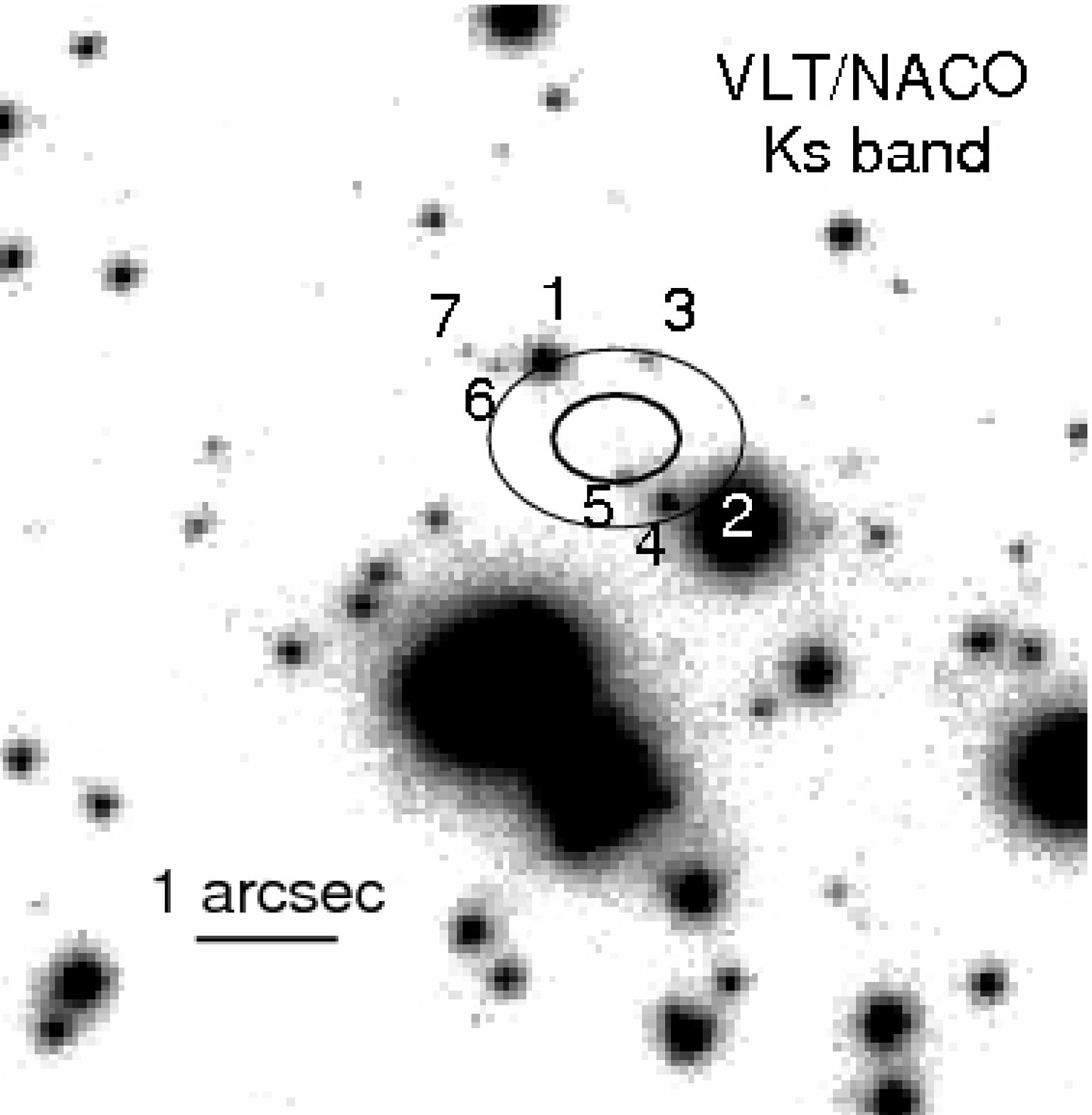}
\caption{Inner part of  the field  of 1E 1613 as
observed  by   the  \vlt$/$\isaac\  ($H$-band),   the  \hst$/$\nicmos\
($160W$)   and   \vlt$/$\naco\    ($K_s$). 
The black box on the \vlt$/$\isaac\ image marks the portion of 
the field shown in the \hst$/$\nicmos\ and \vlt$/$\naco\ images.
North  to the top, East  to the left. In  each case, the
images are the result of the co-addition of repeated integrations (see
Tab. 1--3), with corresponding total integration times of 14000s, 10360
s and 40000 s,  respectively. The inner ellipse corresponds to the 68\% c.l. 
error region, while  
the outer ellipse corresponds to the 99\% c.l. region.
The semiaxes of the 68\% and 99\% ellipses were computed by summing
(in quadrature) the uncertainty on the IR image astrometric
calibration (see \S3.4) to the uncertainty on the Chandra coordinates
(\S2), then multiplying the resulting values by $(-2log(1-0.68))^{(1/2)}$
and by $(-2log(1-0.99))^{(1/2)}$, respectively \cite[see e.g.][]{stat}. 
Candidate  counterparts  detected
within or close to the X-ray  error region are numbered. Object \#1 is
the originally proposed counterpart of \cite{sanwal02}. Objects 3$\div$7 
are resolved only in the \hst$/$\nicmos\ and \vlt$/$\naco\ images.
\label{charts}}
\end{figure*}

Fig. ~\ref{charts} shows the deepest \vlt\ and \hst\ images of the field  
of 1E 1613 with the  computed \chan\  position  overlayed 
(see also the caption to Fig. ~\ref{charts}).  Seven
objects (labelled  in the figure) are  detected in the vicinity of the 
X-ray position; three of them are consistent with the 99\% c.l. error region.
In all cases, the
objects' profiles  are point-like and consistent  with the instruments
PSFs.   We note that  the originally  proposed counterpart  object \#1
\citep{sanwal02} now is only marginally consistent  (at $\sim 3 \sigma$
c.l.) with  the position of
1E 1613.
   Due to  the field crowding,  only the  two brightest
objects (\#1 and 2) are  detected in the lower resolution \vlt/\isaac\
image (Fig.  ~\ref{charts}, left), while the faintest ones (\#3-7) are
detected  in  the  higher  resolution  \hst/\nicmos\  and  \vlt/\naco\
images.  Objects \#1  and 2 are detected also  in the \vlt/\isaac\ $J$
and $K_s$-band  images.  All objects are detected  in the \hst/\nicmos\
$160W$  and  $205W$-bands, while  only  objects  \#1  and 2  are  also
detected in  the $110W$ band,  although the former is  only marginally
detected. This  is likely  due to the  short exposure time  (see Table
\ref{NICMOSdata}) and to the fact  that the $110W$-band is slightly bluer
than  the \isaac\  $J$-band.   The  fact that  objects  \#3-7 are  all
detected in  the $160W$  and $205W$  bands but not  in the  $110W$ one
suggests that they are very red and heavily absorbed.
To be conservative, we will consider all of the seven sources in our 
investigation.

\section{Data analysis}

Due to the field crowding and  to the faintess of most candidates, for
all  observations  magnitudes  were  computed through  PSF  photometry
which,  in this  case, yields  more accurate  results than
standard   aperture  photometry.    We  note   that  for   the  \naco\
observations the  accuracy of  the PSF photometry  is affected  by the
quality  of  the AO  correction,  which was  not  optimal  due to  the
relatively  large  offset  of  the  guide  star  and  to  the  varying
atmospheric  conditions  (\S3.1).  Since  the  \naco\  PSF is  largely
oversampled,  to increase  the signal--to--noise  ratio  \naco\ images
have been  resampled with  a $3\times3$ pixels  window using  the {\em
swarp} program  (Bertin E., Terapix Project).  For  the PSF photometry
we used the Stetson (1992, 1994) suite of programs \dao, following the
procedures described in Zaggia et al. (1997).
To improve  and maximize the object  detection we used  as a reference
the co-added  and deeper  \vlt/\isaac\ $H$, \hst/\nicmos\  $160W$, and
\vlt/\naco\  $K_s$-band  images  (see  Fig.  ~\ref{charts})  for  each
telescope/instrument dataset to create  a master list of objects which
we  registered on  the single  images and  used as  a mask  for object
detection. For each single image,  the model PSF was then computed by
fitting the profile of a  number of bright, but non-saturated, reference
objects. Such model PSF was used  to  measure  object fluxes  at  the
reference  positions.  Photometry  calibration was  applied  using the
zeropoints provided  by the \vlt\ and \hst\  data reduction pipelines.
Since  the  \isaac\ and  \naco\  zeropoints  are  by default  computed
through aperture  photometry, the  aperture correction was  applied to
the \dao\  magnitudes.  For the \vlt\  observation, airmass correction
was applied using the atmospheric extinction coefficients measured for
the Paranal Observatory \citep{patat04}.

\subsection{Short term variability}

To search  for short-term variability from  the candidate counterparts
we  started from  the \vlt/\naco\  dataset  which is the  only  one to
provide both a complete coverage and an accurate phase sampling of the
6.67  hours period of  the X-ray source  over at least  two cycles
(see Table \ref{NACOdata}).   Using as a mask the  master list created
from the  co-added $K_s$-band image we  have run \dao\  to compute the
PSF photometry on the single images.
The derived  single object catalogues  were then matched  and compared
using the \dao\ routine \all.   To avoid systematic offsets induced by
night-to-night  zeropoint fluctuations  the photometry  of  the second
night was renormalized to the first one.

\begin{figure}
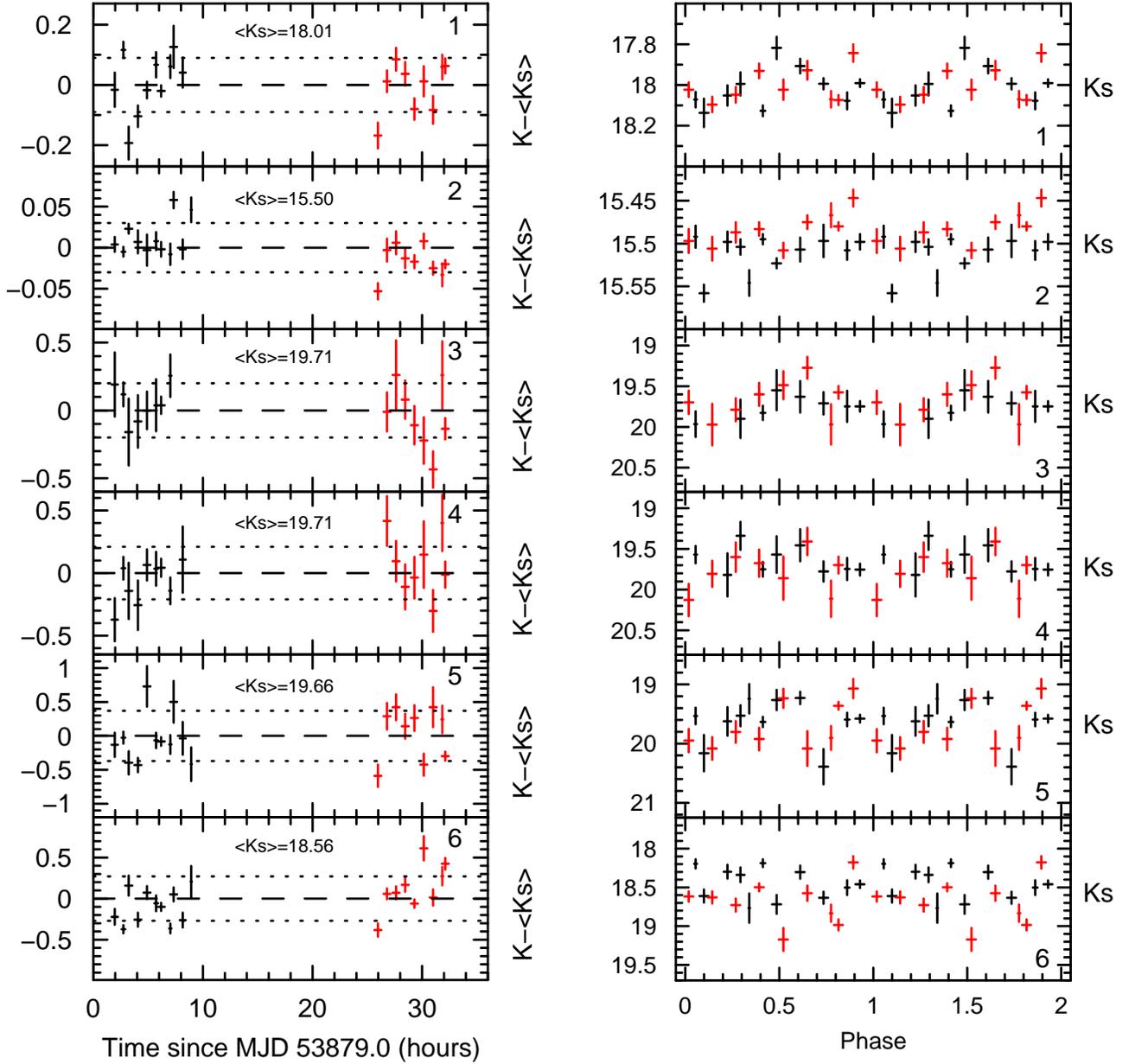

\includegraphics[angle=-90,width=9cm]{f2a.eps}
\includegraphics[angle=-90,width=8cm]{f2b.eps}
\caption{(left) 
Top to bottom: \vlt$/$\naco\ $K_s$-band lightcurves for the candidate
  counterparts \#1-6. Difference with respect to the average 
magnitude is plotted as a function of the time. Horizontal dotted lines mark 
the r.m.s. variability for each source. (right) Top to bottom: folded \vlt$/$\naco\ $K_s$-band lightcurves for the
  candidate counterparts \#1-6. Black and red points represent
  flux measurements performed in the first and in the second night, 
respectively. Two phase intervals are plotted for clarity. 
Error bars 
in both panels account for statistical uncertainties only. 
The r.m.s. variability for each source - plotted in the left panel 
only - is representative of the random errors affecting our 
measurements (see text). 
\label{naco_lc}}
\end{figure}

Magnitude  differences with respect to  the average value are plotted  in
Fig.~\ref{naco_lc} (left panel)  for all candidate counterparts
for the two   consecutive  nights  (May   23rd  and   24th).
Object \#7  is not included since  it is not
detected  in the  single  images but  only  in the  co-added one  (see
Fig. ~\ref{charts}).  
Next, for each measurement, we computed the corresponding phase with respect 
to the 6.67  hours X-ray period,  assuming phase 0  to be at  MJD 53879.0
and we folded the second night on
the first  one.  The folded light curves are  plotted in
Fig.~\ref{naco_lc} (right panel).  

A large scatter of the flux measurements is apparent in both panels of
Fig.~\ref{naco_lc}, where error bars account
for statistical errors only.
Such a variability is generally erratic and not correlated with phase
(Fig.~\ref{naco_lc}). Indeed, visual inspection of the light curves of object \#1
and \#3 would suggest a nearly sinusoidal variation, which, however, 
does not pass statistical tests. 
Considering source \#1, a simple constant fails to reproduce the folded
light curve ($\chi^2_{\nu}=3.9$, 18 d.o.f.). Adding a $sin$ function does not yield a better
description ($\chi^2_{\nu}=3.6$, 16 d.o.f.; such improvement has a 20\%
chance occurrence probability); adding a second harmonic does not improve
the situation ($\chi^2_{\nu}=4.0$, 14 d.o.f.). 
Focusing on source \#3, a constant fit yields $\chi^2_{\nu}=1.5$ (15 d.o.f.),
while adding a $sin$ function yields $\chi^2_{\nu}=1.0$ (13 d.o.f.).
Such an improvement has a chance occurrence probability of 3\%, which is definitely
too high to claim for evidence of modulation.

Thus, none of the candidate counterparts show evidence for periodic modulation.

In order to estimate upper limits on short-term variability,
a careful discussion of errors is required. 
Indeed, the apparent flux variations exceed the expected poisson fluctuations
among different measurements. A steady flux model is formally
not consistent with most of the observed light curves 
(as seen above, e.g., for object 1 it yields $\chi^2 _{\nu}=3.9$, 18 d.o.f.). 

\begin{figure}
\includegraphics[angle=0,width=8cm]{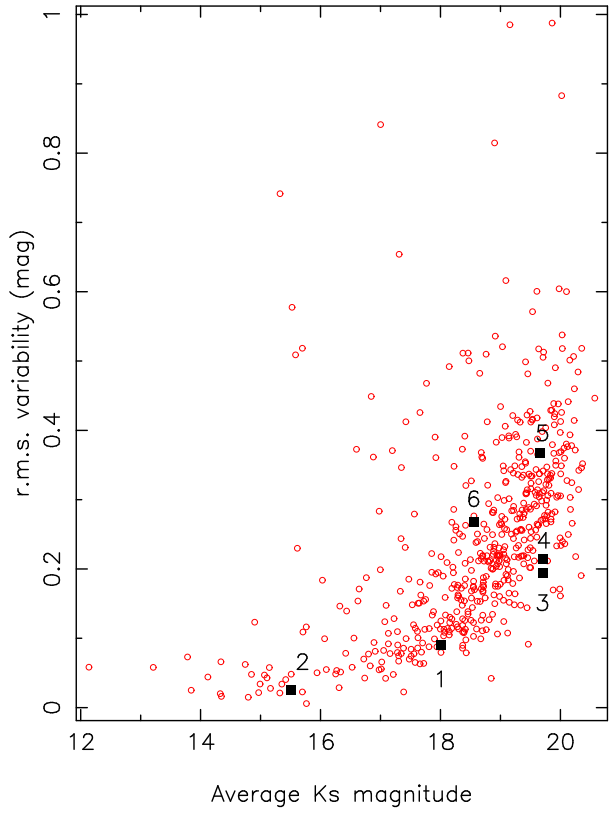}
\caption{The r.m.s. variability for all sources detected in the \vlt$/$\naco\
field is plotted as a function of the source's average magnitude.  
A larger  variability for fainter sources is apparent.
Black
squares represent the possible candidate counterparts (sources 1-6, see text).
\label{fig:rms}}
\end{figure}

Such a large scatter suggests that  our relative photometry
measurements are contaminated by random errors induced e.g.,
by  fluctuations in  the  atmospheric conditions,  sky background  and
seeing.  Other sources  of errors are the  variations in the AO
correction  to  the  image  PSF,  which depend  both  on  the  objects
positions in the  instrument field of view and  on the correct guiding
of the reference star.  Furthermore, errors are also induced by
the PSF fitting and  background subtraction procedures.  Other  errors
induced  by the  data reduction  process  and/or by  glitches in  the
detector performance should also to be taken into  account. 
It is also  interesting to note that the brightest source 
candidate (object  \#2, $K_s$=15.5) shows a much  smaller r.m.s.  with
respect to  the other, much  fainter ones ($K_s> 18$), implying
that such errors are larger for sources with a lower signal to noise ratio. 
Since  it  would  be  extremely  difficult  to  formally
quantify  all the  above  effects  on the  photometry  of each  single
object, we decided to use  an empirical approach.  The results are
shown in Fig.~\ref{fig:rms} where, for  all objects detected in the field,
we have  plotted the flux  variation r.m.s. as  a function of
the  average object's  flux.  As already  hinted in
Fig.~\ref{naco_lc},  flux  measurements for  fainter
sources are  more scattered.  In particular, the  flux variation r.m.s
of our candidate counterparts are fully consistent with those of field
objects of comparable brightness.  Thus,  we conclude that none of the
candidate  counterparts  shows evidence  for a significant short-term
variability, nor  of  a significant 6.67 hour  modulation.
Indeed, the
measured flux  variation r.m.s. can be interpreted as a
$1\sigma$  upper limit  to any
possible  variability.  Such upper limits,  together with  the
time-averaged magnitudes, are summarized in Table \ref{tab:rms}.

For completeness, we have  analyzed the other available datasets which
span  different epochs and  sample different  source states.   We have
repeated  our analysis  using  the \vlt/\isaac\  $H$-band dataset  (see
Table  \ref{ISAACdata}) which,  unfortunately, provides  repeated flux
measurements only for the two brightest candidates (objects \#1 and 2)
with non-continuous  coverage  and  non-unifourm
sampling of the 6.67 hours X-ray period.  We found that the apparent flux
variations are compatible  with the measured  r.m.s., suggesting
that, also  in this case, random  errors induced by  fluctuations in the
atmospheric  conditions,  sky  background  and  seeing,  dominate.  
As before, we
assumed  the   measured  flux   variation  r.m.s.  as   the  $1\sigma$
variability upper limit (see  Table \ref{tab:rms}).

Finally,  we have repeated  our analysis  using the  \hst/\nicmos\ $160W$
dataset  which  provides  repeated,
atmosphere-free   flux  measurements 
although with non-continuous coverage
and non-uniform sampling  of the 6.67  hours X-ray period  (see Table
\ref{NICMOSdata}). 
Since the \hst\
observations  have been  split in  two different  orbits  separated by
about 60 days,  we first searched for variability  
in each visit.
Once more, we could
not  find  statistically  significant  evidence for  variability.   In
particular, we  found none  for object \#7,  i.e. the faintest  of the
candidates  and  the  only  one  for which  the  repeated  \vlt/\naco\
observations   did    not   yield   time    variability   information.
While the uncertainty on the source period prevents
folding of the light curves derived from different visits,
by comparing the photometry  of the two visits we found
that  the  fluxes measured  from  the  second  one are  systematically
fainter  by  $\sim 0.1-0.2$  magnitudes. Such an effect is most probably due
to the use of calibration frames non contemporaneous to the observations
(unfortunately, a single set of dark and flatfield frames is available 
for the two visits) and, possibly, to a 
non-optimal  correction for  the ``pedestal''  (see  NICMOS Instrument
Handbook). 
The derived variability  upper limits (see Table \ref{tab:rms})
are  thus less constraining  than  those derived  from  the single  visits,
although, as  expected, somewhat tighter than those  derived from the
analysis of the \vlt/\naco\ and \isaac\ datasets.

\subsection{Long term variability}

We have used the whole dataset to search for indications of long term,
IR variability on time scale of years,  possibly  associated to the evolution
of the  X-ray source. As a reference, we have
used  the flux  measurements  in the  closest  passbands, i.e.   those
obtained  from the \vlt/\isaac\  $K_s$-band, \hst/\nicmos\  $205W$ and
\vlt/\naco\ $K_s$-band observations.  
Since we did not find any evidence for variability within the dataset of
each instrument, we have generated, for each dataset, time-averaged images.
Unfortunately, a direct comparison is made difficult by the use of different
instruments and filters.
Passband transformations  between the \hst\ filters  and the Johnson's
ones can  be
computed    
using    the     {\em synphot}\footnote{http://stsdas.stsci.edu/cgi-bin/gethelp.cgi?synphot.sys}
 package of the Space Telescope Science Data
Analysis Software (STSDAS). 
However,  the
results are affected  by the uncertainty on the  objects spectral type
which is a free parameter  of the transformation equation.  Since 
our spectral classification  of the  candidate counterparts
relies mostly  on one colour, with only  objects \#1 and  2 also
detected in the \vlt/\isaac\ $J$-band and in the \hst/\nicmos\ $110W$,
we estimate  that a straight passband tranformation  will introduce an
unknown uncertainty in our flux estimates.

To solve the problem, as well as to  account for other sources
of systematics,  we cross-matched the object  catalogues obtained from
the   \hst/\nicmos\  $205W$-band   and   the  \vlt/\isaac\   $K_s$-band
observations (202 objects in common)  and we computed  the correlation
between the magnitudes measured in  the two filters.  The r.m.s of the
fit is $\sim$  0.05 magnitudes (using 105 sources with $H<19.5$ after
2$\sigma$ clipping), i.e. of the same  order of our statystical
photometric  errors. 
We  repeated the  same procedure  to  compute the
transformation  between the  \vlt/\naco\ and  \isaac\  $K_s$-bands and
again we  obtained an r.m.s. of  $\sim$ 0.05 magnitudes
(using 47 sources with $K<17$, after $2.5\sigma$ clipping).   For all our
candidates   we   then  applied   such  empyrical   passband
transformations in  order to  have all flux  measurements consistently
referred to the $K_s$-band. 

We also checked the correlation for the bulk of fainter
sources, resolved only in the sharp \hst/\nicmos\ and \vlt/\naco\
images. The scatter is found to be somewhat higher at faint fluxes.
The observed r.m.s. increases from 0.08 mag (for 19 sources in the $K_s$ 13-15.5
range) to $\sim0.4$ mag (for 243 sources in the $K_s$ 18-20.5 mag range).  
The larger r.m.s. for fainter objects 
points to effects
related e.g. to passband transformation, airmass corrections, as well as to 
the effects discussed in \S 4.1. 

Fig.~\ref{multi_lc}  shows the derived  long term  $K_s$-band lightcurve
for all our candidates  compared with the \chan/\xmm\ X-ray lightcurve
obtained over the  same time span.  Note that,  since only objects \#1
and 2 were detected in  the \vlt/\isaac\ observations, for the fainter
candidates the lightcurve is based on the \hst/\nicmos\ and \vlt/\naco\
points only. 
While the X-ray source was 
continuously fading, for most candidates there  is no indication of
IR variability  on the year time  scale.  
Object \#6  
apparently decreases by  $\sim  1$ magnitude between
August 2002 and May 2006.  However,  we note that this object falls in
the   PSF  wings   of  the   much  brighter   object  \#2   (see  Fig.
~\ref{charts}),  which might  have affected  our photometry. 
A  possible
$\approx0.7$  magnitudes flux decrease  is also observed  for objects
\#4 and 5 (with the latter being the only candidate to
fall within the  68\% c.l. X-ray error circle).  
However,  in view of
the larger uncertainties for fainter sources --
as apparent from the scatter in the
   \hst/\nicmos\ - to - \vlt/\naco\ correlation --  such evidences
for variability should be taken with  caution.



\subsection{Color analysis}

\begin{figure*}
\includegraphics[angle=0,width=10cm]{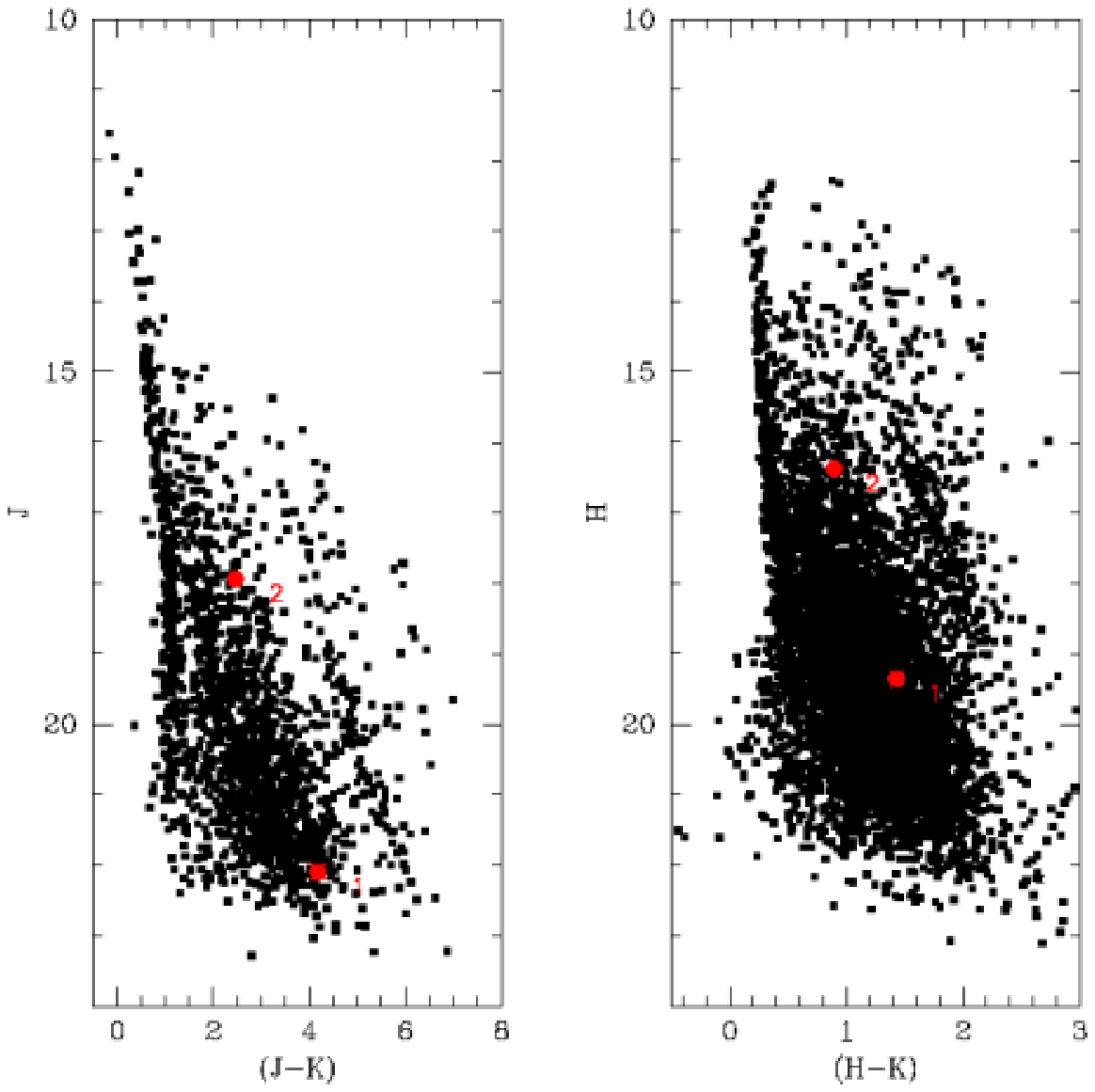} \\
\includegraphics[angle=0,width=10cm]{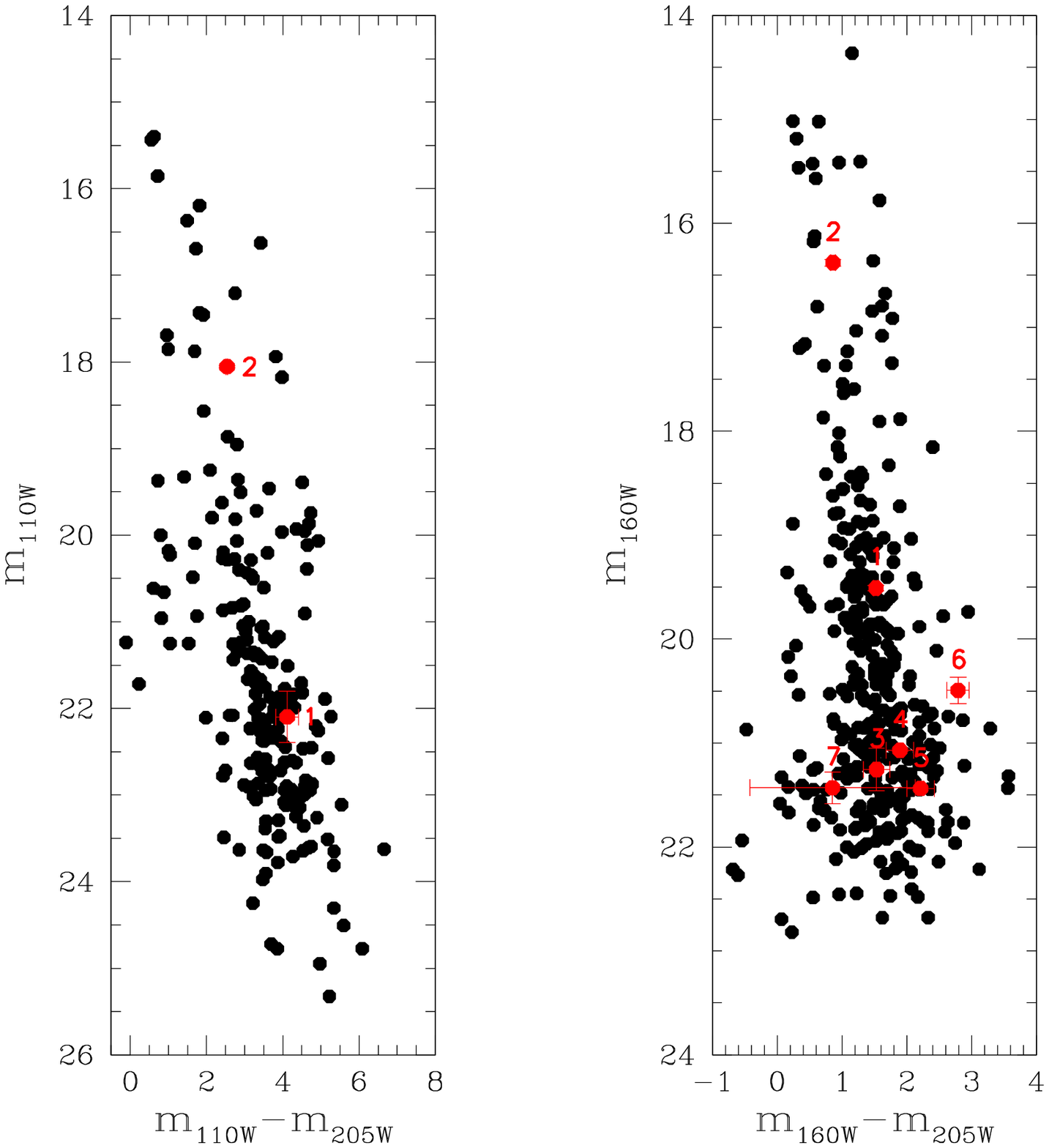}
\caption{($J,J-K_s$) and ($H,H-K_s$) CMDs of the  field of 1E 1613 obtained from \vlt/\isaac\ (upper pair) and the \hst/\nicmos\ (lower pair) observations. The locations of the candidate counterparts (see Fig.~\ref{charts}) are plotted in red. 
\label{cmd}}
\end{figure*}

We used  the available multi-band  information to derive clues  on the
nature of  the candidate counterparts.  The single  band, single epoch
photometry  catalogues derived  with \dao\  were matched  by  \all\ to
create   the   multi-band   catalogues   for  the   \vlt/\isaac\   and
\hst/\nicmos\ datasets.  Single epoch multi-band catalogs were finally
merged  and   magnitudes  averaged.   We  note  that   since  no  time
variability  has  been found  in  the  \vlt/\isaac\ and  \hst/\nicmos\
observations  (see \S  4.1), the  use of  average magnitudes  for each
dataset does  not affect our  color analysis.  To make  the comparison
between  the derived  \vlt/\isaac\  and \hst/\nicmos\  color-magnitude
diagrams (CMDs) consistent, \nicmos\ $110W$ and $160W$ magnitudes have
been transformed to the \isaac\  $J$ and $H$ bandpasses using the same
approach applied  in the  previous section.  The  available multi-band
photometry of  our candidates is summarized  in Table \ref{multiphot}.
The results of our color  analysis are shown in Fig.  ~\ref{cmd} where
we  plot  the  candidates  photometry  on  the  \vlt/\isaac\  and  the
\hst/\nicmos\  ($J,J-K_s$)  and  ($H,H-K_s$)  CMDs  built  from  the
photometry of  the field  stars. As seen,  nearly all  candidates have
colors consistent with the bulk  of the field stellar population.  The
only possible exceptions are object \#7, whose colour determination is
however affected by the large  photometry errors, and object \#6 which
seems to be slightly redder with respect to the CMD sequence which has
an average $H-K_s \sim  2$.  However, this apparently peculiar deviation
could be partly due to the lower statistics of the \hst/\nicmos\
CMD, where  the redder  part is poorly  sampled. Indeed,  objects with
extreme $H-K_s$ appear less unusual in the much denser \vlt/\isaac\ CMD.
Furthermore, we warn here that object  \#6, as well as perhaps \#4 and
5, might  be variable on the  long time scale (see  \S4.2).  Thus, its
location in the CMD might not be fully representative.

\subsection{Deep imaging}

No other candidates have been  identified in our deep IR imaging within
or  close to the  X-ray position  apart from  those indicated  in Fig.
~\ref{charts}.  We  have used  our deepest images  of the  field, i.e.
those  obtained  from the  co-addition  of  the repeated  \vlt/\isaac\
$H$-band,  the \hst/\nicmos\  $160W$, and  the  \vlt/\naco\ $K_s$-band
observations,  to set  constraining upper  limits  on the  flux of  an
hypotethically undetected CCO counterpart.  We obtained $H\sim 23$ and
$K  \sim  22.1$  (both  at  $3 \sigma$)  from  the  \hst/\nicmos\  and
\vlt/\naco\  observations, respectively.   These values  represent the
deepest upper  limits obtained so far  for this source.   

\section{Discussion}

Although our   comprehensive   study   (astrometry,   variability,   multi-band
photometry) of  the potential CCO  counterparts did not single  out an
high confidence  candidate,  position-wise, object \#5 (inside the 68\% c.l. region),  
objects \#3 and \#6 (inside the 99\% c.l. region), as well as object \#1
(marginally consistent, at the $\sim3\sigma$ level) cannot be ruled out.
On the other hand, objects 
\#2, \#4 and \#7 may be disregarded.
The field is very crowded, with a source density in the combined \vlt/\naco\ 
$K_s$ image of $>1.1$ objects per square arcsec at the sensitivity 
limit of $K_s \sim 22.1$.  
Before discussing the implications of our results on possible pictures
for 1E 1613, we note that
none of the possible candidates stands  out for peculiar
colors  with  respect  to  the field very red  stellar  population. The
average  H-K$\sim2$ requires a large interstellar reddening,
consistent with  A$_V \sim 20-25$. 
Such a reddening is much larger than the value of  A$_V \sim 3.3-6.6$
expected  at the distance of the X-ray source,
according to
the measured N$_H$ of 1E 1613 \citep{deluca06}, to the results of neutral H study towards
RCW103 \citep{reynoso04} as well as to spectrophotometry of the SNR
\citep[e.g. ][]{leibowitz83}.

Thus, if one of the plausible candidates  is indeed
physically associated with 1E 1613, it must
have very peculiar, red intrinsic colors. 

In the frame of the binary system scenario
for 1E 1613 \citep{deluca06}, in principle, one could expect the companion
star to be significantly different from a main sequence 
star of comparable mass. This could be the result of an early phase 
of irradiation by photons and charged particles from the newborn neutron
star, which could have left the companion away from thermal equilibrium
(Kelvin-Helmoltz time scale to thermal relaxation
would be much larger than the age of the system).
Tidal interaction along a very eccentric orbit could also play some role.
However, the hypothesis that any of the possible candidates be the companion of 1E 1613
is very unlikely. The observed colors and magnitudes
would require an unrealistically low temperature for the star
($\sim1000 - 1500$ K), implying (at a distance of 3.3 kpc and for A$_V=6.6$), 
a photospheric radius of  (1-2)$\times10^{11}$ cm, exceeding the 
Roche lobe dimension ($\sim4\times10^{10}$ cm, assuming a 1.4 M$_\odot$ neutron star, 
a 0.5 M$_\odot$ companion
and an orbital period of 6.67 hours) and comparable to the system orbital 
separation ($\sim1.5\times10^{11}$ cm under the same assumptions).

Within the isolated magnetar scenario
\citep{deluca06}, a fallback disc is required
in order to quench the neutron star rotation to a period of 6.67 hours in
$\sim2000$ yr. Could one of the possible candidates
be the fallback disc itself? Evidence for a debris disc surrounding
the anomalous X-ray pulsar 4U 0142+61, recently obtained \citep{wang06}, 
make such an hypothesis not unrealistic \citep[a faint IR source
at the position of the CCO in Vela Jr. could also be related to a debris disc 
surrounding the compact object, see][]{mignani07b}. While the physics
of the possible disc surrounding AXP 4U 0142+61 is not understood 
\citep[passive or viscous? ][]{wang06,ertan07}, we note that 
the colors of our candidates are similar to the case of 4U 0142+61,
but their F$_{Ks}$/F$_X$
ratio (in the range 0.7-2$\times10^{-3}$
at the epoch of the NACO observations) is about one order 
of magnitude larger. The upper limits to the far IR emission set by
\citet{wang07} are not constraining.

Unfortunately, mostly because of the faintness of the IR sources, results of the
temporal analysis could not offer conclusive clues.
The upper limits we could set
do not put stringent constraints. They could be consistent
with a binary scenario, since any orbital modulation would obviously
depend on the inclination of the system with respect to the plane of the sky.
On the other side, the isolated neutron star scenario
does not allow for firm predictions about a possible IR periodicity.

We also tried to exploit the source long-term fading seen in X-rays
between the NICMOS and the NACO observations looking for long-term
variability to pinpoint the IR counterpart of 1E 1613.
In the binary scenario, IR variability could be due to the reprocessing of
the compact source X-ray radiation by the companion star (and possibly by an
accretion disc, if any). The same could be true for the isolated magnetar
scenario, X-ray reprocessing taking place in a fallback disc.
We stress, yet again, that in both pictures the effect would be strongly
dependent on the geometry of the system with respect to the line of sight.
Indeed, we obtained evidence for a possible $\sim1$ mag fading of one of the candidates,
which 
yields a constant F$_{Ks}$/F$_X$ ratio in case of association to 1E 1613.
However, in view of the scatter seen in the correlation of
NICMOS and NACO photometric measurements for sources of comparable magnitude, such a possible
evidence for IR variability is not strong enough to claim an identification.



With no compelling reasons to associate any of the possible 
candidates to 1E 1613, we explore the implications of a source non-detection down  
to Ks$>$22.1. 

Considering the binary scenario, accounting for 
uncertainties on the distance and reddening, 
such upper limit is only consistent with  an M6-M8 dwarf,
i.e. a very underluminous companion. 
It is rather unlikely that such a small star may power the mechanism
proposed by \citet{deluca06}, 
where accretion
of wind from the companion -- 
modulated along an eccentric orbit -- 
explains the 
phenomenology of the low state. Such a picture would require
an accretion rate of $\sim10^{-13}\,M_{\odot}\,y^{-1}$,
implying a red dwarf wind mass loss 
larger by at least a factor of a few, which seems somewhat too
high for such a small star.
Thus, an alternative process is required to
explain the X-ray phenomenology.
Even the survival to the supernova explosion  of a 
binary system with such an extreme mass ratio 
seems problematic.
It would require an ad-hoc kick for the neutron star
to avoid disruption of the system. 

On the other side, the lack of an IR counterpart fits well within the
isolated magnetar scenario. The upper limit to the ratio
F$_{Ks}$/F$_X<1.5\times10^{-4}$ 
(at the epoch of the NACO observation) is fully consistent with the values
observed for all magnetars identified in the IR
\citep[see e.g.][]{fesen06}, including 4U 0142+61 
\citep{wang06}.
In the isolated magnetar scenario,
we cannot rule out the possibility that one of the candidate sources be 
the residual disc surrounding 1E 1613.
However, Occam's razor argues against such a conclusion, since
all possible candidates are undistinguishable from normal, background stars.

\begin{figure}
\includegraphics[angle=0,width=7cm]{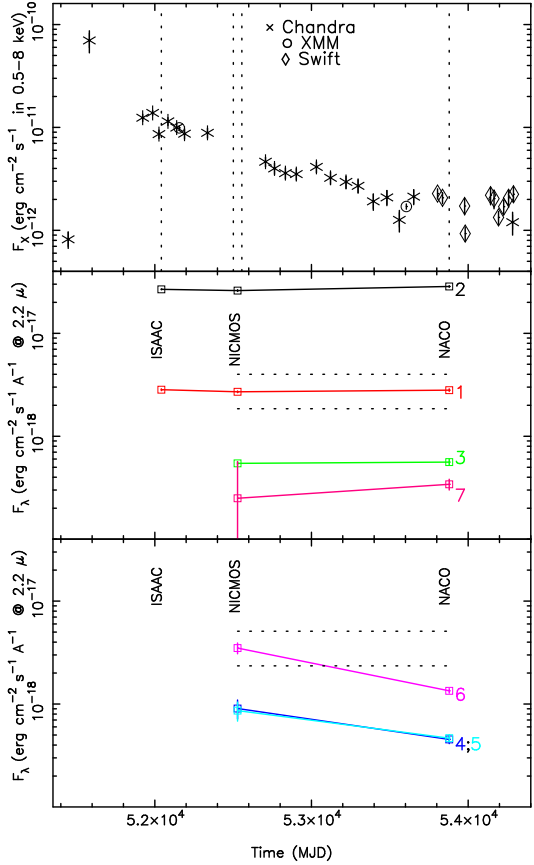}
\caption{Upper panel: X-ray lightcurve of 1E 1613  in the time range
1999, September to 2007, July, 
updated from
  \citet{deluca06}. Data have been
collected with the Chandra/ACIS, Chandra/HRC, XMM-Newton/EPIC and Swift/XRT
instruments.  
Swift/XRT as well as Chandra/HRC fluxes have been computed
  assuming the source spectral shape to be the same as observed by XMM-Newton
  in 2005 \citep{deluca06}. Error bars are at $1\sigma$ confidence level. 
 Details on the analysis of the X-ray dataset will be
  reported elsewhere. Vertical dotted lines mark the epochs of the IR $K_s$ observations.
Middle panel: $K_s$-band lightcurves for candidate counterparts
\#1,2,3,7. Error bars
account for $1\sigma$ statistical uncertainty. Horizontal dotted lines
mark the r.m.s. observed in the correlation of NICMOS and NACO
measurements for sources in the K$\sim$18.5-20 magnitude range,
which could be used as an estimate of the overall uncertainty. 
Lower panel: same as middle panel for candidate counterparts \#4-6,
showing a possible flux decrease.
The possible variation for source \#6 is about twice the 
uncertainty in the NICMOS-to-NACO comparison.  
\label{multi_lc}}
\end{figure}

Recent SWIFT monitoring of 1E 1613 shows that the X-ray source continued to fade in 
the 2006-2007 time interval 
and that its flux is approaching the 1999, pre-outburst level (see Fig.~\ref{multi_lc}). 
Historical observations point to a large flux variability
over a $\sim10$ yr time scale \citep{gotthelf99,deluca06}, so that a rebrightening
is likely to occur. A factor $\sim100$ increase in the X-ray luminosity
(as in the 1999-2000 outburst) would yield a dramatic change in the irradiation
of any possible object/structure (companion star/fallback disc)
linked to 1E 1613. 
Thus, if a new outburst from 1E 1613 will occur,
a fast follow-up in the IR will be crucial in order to 
conclusively address the issue of the IR counterpart for 1E 1613
and to shed light on its nature. The images described in this work
will be a reference to search for  variability
of the counterpart. 


\acknowledgments
This research is partially supported by the Italian Space Agency (contract
ASI I/011/07/0 in support to the Swift mission).

\clearpage

\begin{table}
\begin{center}
  \caption{Summary of  the \vlt/\naco\ observations of  the field of 1E 1613, with the observing epochs, the observations start time (UT), the filter, the exposure times, the average seeing and airmass  values.eld. }
\begin{tabular}{cccccc} \\ \hline
 yyyy-mm-dd    & Time (UT) & Filter & T (s) & Seeing (``) & Airmass	\\ \hline
2006-05-24  &  00:49:53    &     Ks    &   2280    &   0.84  &   1.57\\
            &  01:57:36    &     Ks    &   2040    &   0.64  &   1.36\\
            &  02:45:14    &     Ks    &   1200    &   1.10  &   1.23\\
            &  03:14:21    &     Ks    &   2280    &   1.02  &   1.18\\
            &  04:04:24    &     Ks    &   2280    &   0.91  &   1.13\\
            &  04:54:17    &     Ks    &   2280    &   0.81  &   1.12\\
            &  05:44:25    &     Ks    &   1200    &   0.80  &   1.14\\
            &  06:11:50    &     Ks    &   2280    &   0.90  &   1.16\\
            &  07:02:02    &     Ks    &    720    &   1.06  &   1.24\\
            &  07:19:58    &     Ks    &   2280    &   0.95  &   1.28\\
            &  08:09:50    &     Ks    &   1920    &   1.00  &   1.44\\
            &  08:56:06    &     Ks    &    360    &   0.89  &   1.67\\
2006-05-25  &  00:55:12    &     Ks    &   2280    &   0.61  &   1.54\\
            &  01:57:43    &     Ks    &   2280    &   0.66  &   1.34\\
            &  02:47:40    &     Ks    &   2280    &   0.60  &   1.22\\
            &  03:37:25    &     Ks    &   2280    &   0.65  &   1.15\\
            &  04:27:26    &     Ks    &   2280    &   0.64  &   1.12\\
            &  05:17:22    &     Ks    &   2280    &   0.62  &   1.12\\
            &  06:08:53    &     Ks    &   2280    &   1.00  &   1.16\\
            &  06:59:57    &     Ks    &   2280    &   0.96  &   1.24\\
            &  07:50:24    &     Ks    &    600    &   0.72  &   1.38\\
            &  08:06:16    &     Ks    &   2280    &   1.00  &   1.44\\ \hline
\end{tabular}
\label{NACOdata}
\end{center}
\end{table}

\begin{table}
\begin{center}
  \caption{Summary of  the \vlt/\isaac\ observations of  the field of 1E 1613, with the observing epochs, the observations start time (UT), the filter, the exposure times, the average seeing and airmass  values. }
\begin{tabular}{cccrcc} \\ \hline
yyyy-mm-dd     & Time (UT) & Filter & T (s) & Seeing (``) & Airmass	\\ \hline
2001-04-10  &  07:33:07    &      J    &   2000    &   0.68  &   1.12 \\ 
            &  07:33:07    &      J    &   2000    &   0.68  &   1.12  \\
2001-05-12  &  06:04:33    &     Ks    &   2000    &   0.95  &   1.13  \\
2001-05-13  &  04:42:57    &     Ks    &   2000    &   1.00  &   1.13  \\
            &  05:39:35    &     Ks    &   2000    &   1.41  &   1.12  \\
2001-06-06  &  05:55:32    &      H    &   1000    &   0.64  &   1.28  \\
            &  06:46:57    &      H    &    200    &   1.04  &   1.34  \\
2001-07-23  &  01:48:12    &      H    &   1000    &   1.03  &   1.14  \\
            &  02:20:50    &      H    &   1000    &   0.98  &   1.18  \\
            &  02:50:03    &      H    &   1000    &   1.02  &   1.22  \\
2001-07-29  &  00:17:16    &      H    &   1000    &   0.73  &   1.12  \\
            &  02:45:57    &      H    &    950    &   0.70  &   1.26  \\
            &  03:14:54    &      H    &   1000    &   0.84  &   1.33  \\
            &  03:43:08    &      H    &   1000    &   0.70  &   1.43  \\
2001-07-30  &  00:27:13    &      H    &   1000    &   1.00  &   1.12  \\
            &  01:04:19    &      H    &   1000    &   1.00  &   1.13  \\
            &  01:32:56    &      H    &   1000    &   1.00  &   1.15  \\
            &  02:01:25    &      H    &   1000    &   1.00  &   1.18 \\ \hline
\end{tabular}
\label{ISAACdata}
\end{center}
\end{table}

\begin{table}
\begin{center}
  \caption{Summary of  the \hst/\nicmos\ observations of  the field of 1E 1613, with the observing epochs, the observations start time (UT), the filter and the exposure times. }
\begin{tabular}{cccr} \\ \hline
yyyy-mm-dd     & Time (UT) & Filter & T (s) \\ \hline
2002-08-15     & 02:25:38 & $160W$ & 2590                      \\
               & 04:10:28 & $160W$ & 2590                    \\
               & 06:05:15 & $160W$ & 2590                     \\
               & 09:09:37 & $160W$ & 2590                     \\
               & 11:12.32 & $110W$ & 935                      \\
               & 12:40:18 & $205W$ & 1007                     \\ 
2002-10-08     & 08:50:00 & $160W$ & 2590                     \\
               & 10:27:32 & $160W$ & 2590                     \\
	       & 12:04:01 & $160W$ & 2590                     \\
	       & 13:40:14 & $160W$ & 2590                     \\
               & 15:19:28 & $110W$ & 935                      \\
               & 15:42:17 & $205W$ & 1007                     \\ \hline
\end{tabular}
\label{NICMOSdata}
\end{center}
\end{table}

\begin{table}
\begin{center}
  \caption{Time-averaged magnitudes and associated r.m.s. (in parenthesis) for all candidate counterparts. Different columns refer to different telescope/instrument/filter combinations.  }
\begin{tabular}{ccccc} \\ \hline
 Id. &   \vlt/\naco\ & \vlt/\isaac\  & \hst/\nicmos\  \\   
     &    ($K_s$)    & ($H$)         & ($160W$) \\ \hline      
 1   & 18.01 (0.09)& 19.082 (0.29) & 19.51  (0.09)\\
 2   & 15.50 (0.03)& 16.411 (0.03) & 16.38  (0.02)\\
 3   & 19.71 (0.20)&   -           & 21.25  (0.08)\\
 4   & 19.71 (0.21)&   -           & 21.07  (0.04)\\
 5   & 19.66 (0.37)&   -           & 21.44  (0.16)\\
 6   & 18.56 (0.27)&   -           & 20.49  (0.16)\\ 
 7   &    -          &   -         & 21.43 (0.07) \\\hline
\end{tabular}
\label{tab:rms}
\end{center}
\end{table}

\begin{table}
\begin{center}
  \caption{Multi-band magnitudes for all the candidate counterparts. 
Values are computed on the average images. To allow for an easier comparison
between different measurements, \hst\ magnitudes have been renormalized to the 
Johnson's system (see text). Quoted uncertainties include statistical errors
only.}
\begin{tabular}{c|ccc|ccc|c} \\ \hline
ID &\multicolumn{3}{c}{\isaac} & \multicolumn{3}{c}{\nicmos} & \multicolumn{1}{c}{\naco} \\ 
   & J & H & K & J& H & K & K \\ \hline
\hline
1  &  22.10$\pm$0.10 &  19.36$\pm$0.03 &  17.93$\pm$0.02 & 22.10$\pm$0.30 & 19.50$\pm$0.02 & 17.98$\pm$0.02 & 17.94$\pm$0.03\\
2  &  17.94$\pm$0.01 &  16.38$\pm$0.01 &  15.49$\pm$0.01 & 18.05$\pm$0.01 & 16.38$\pm$0.03 & 15.52$\pm$0.02 & 15.42$\pm$0.02\\
3  &  -              &  -              &  -              & - & 21.25$\pm$0.20 & 19.72$\pm$0.04 & 19.69$\pm$0.07\\
4  &  -              &  -              &  -              & - & 21.07$\pm$0.01 & 19.17$\pm$0.21 & 19.92$\pm$0.09\\
5  &  -              &  -              &  -              & - & 21.43$\pm$0.01 & 19.22$\pm$0.21 & 19.89$\pm$0.08\\
6  &  -              &  -              &  -              & - & 20.49$\pm$0.17 & 17.70$\pm$0.12 & 18.74$\pm$0.05\\
7  &  -              &  -              &  -              & - & 21.43$\pm$0.15 & 20.57$\pm$1.27 & 20.23$\pm$0.12\\ \hline
\end{tabular}
\label{multiphot}
\end{center}
\end{table}


\clearpage

\end{document}